\documentclass[preprint]{aastex}
\usepackage{lscape,amsmath}
\begin{document}

\title{Modeling Tracers of Young Stellar Population Age in Star-Forming Galaxies}
\author{Emily M. Levesque\altaffilmark{1}}
\affil{CASA, Department of Astrophysical and Planetary Sciences, University of Colorado 389-UCB, Boulder, CO 80309}
\author{Claus Leitherer}
\affil{Space Telescope Science Institute, 3700 San Martin Drive, Baltimore, MD 21218, USA}
\altaffiltext{1}{Hubble Fellow; \texttt{Emily.Levesque@colorado.edu}}
\shortauthors{Levesque \& Leitherer} 

\begin{abstract}
The young stellar population of a star-forming galaxy is the primary engine driving its radiative properties. As a result, the age of a galaxy's youngest generation of stars is critical for a detailed understanding of its star formation history, stellar content, and evolutionary state. Here we present predicted equivalent widths for the H$\beta$, H$\alpha$ and Br$\gamma$ recombination lines as a function of stellar population age. The equivalent widths are produced by the latest generations of stellar evolutionary tracks and the Starburst99 stellar population synthesis code, and are the first to fully account for the combined effects of both nebular emission and continuum absorption produced by the synthetic stellar population. Our grid of model stellar populations spans six metallicities ($0.001 < Z < 0.04$), two treatments of star formation history (a 10$^6$ M$_{\odot}$ instantaneous burst and a continuous star formation rate of 1 M$_{\odot}$ yr$^{-1}$), and two different treatments of initial rotation rate ($v_{\rm rot} = 0.0v_{\rm crit}$ and 0.4$v_{\rm crit}$). We also investigate the effects of varying the initial mass function. Given constraints on galaxy metallicity, our predicted equivalent widths can be applied to observations of star-forming galaxies to approximate the age of their young stellar populations.
\end{abstract}

\section{Introduction}
The radiative properties of a star-forming galaxy are generated by its young stellar population. Young hot massive stars are the primary source of ionizing radiation in these galaxies, powering their emission spectra. Supernova and gamma-ray burst progenitors in these galaxies are members of this same massive star sample. The lower-mass members of this young stellar population will often dominate the galaxy's continuum emission. As a result, the age of the newest generation of stars is a crucial factor in determining the origins of a star-forming galaxy's spectral signature. Well-developed age diagnostics for star-forming galaxies therefore allow us to better understand the nature of the galaxies' overall star formation history, the nature of their stellar content, and their current state of evolution and ionization.

The equivalent widths of the strongest hydrogen recombination lines -- H$\beta$, H$\alpha$, and Br$\gamma$ -- are all expected to be excellent indicators of young stellar population age. The steady decrease of the ionizing photon flux as the youngest and hottest massive stars die will deplete the nebular component of the hydrogen lines, while the underlying absorption will decrease at a much slower rate due to the substantial contribution from cooler lower-mass stars. As a result, the equivalent widths of these features will directly measure the ratio of the young ionizing population over the older non-ionizing population (Stasi\'{n}ska \& Leitherer 1996, Leitherer 2005).

Dottori (1981) was the first to predict that the equivalent width of H$\beta$ (W(H$\beta$)) should decrease with the age of a starburst. Soon after, Copetti et al.\ (1986) modeled a starburst HII region and found that W($H\beta$) decreased almost monotonically with age. The first in-depth analysis of this and other similar age diagnostics was presented in Stasi\'{n}ska \& Leitherer (1996). While the nature of the hydrogen recombination lines is complicated by other parameters such as metallicity, the nature of the initial mass function (IMF), stellar mass loss rates, and older underlying stellar populations, both Copetti et al.\ (1986) and Stasi\'{n}ska \& Leitherer (1996) concluded that the primary determining factor in these line widths was starburst age. Finally, Schaerer \& Vacca (1998) presented a suite of evolutionary synthesis model outputs that illustrated the evolution of W($H\beta$) with age for a variety of different metallicities, clarifying the abundance dependence of this age diagnostic. This metallicity-dependent W($H\beta$)-age relation was recently quantified in Levesque et al.\ (2010a). Based on these theoretical relations, W($H\beta$) has since been used as a direct proxy for starburst age (e.g. Stasi\'{n}ska et al.\ 2001). 

More recently, Fernandes et al.\ (2003) were the first to verify from observations that W(H$\beta$) does indeed decrease with starburst age, combining empirical determinations of starburst age with observations and synthetic galaxy spectra. Mart\'{i}n-Manj\'{o}n et al.\ (2008) further extended these age diagnostics to more complex star formation histories; while W(H$\beta$) is a reliable gauge of a single coeval burst's stellar population, most realistic galaxy models much account for multiple star formation episodes. They concluded that even when modeling multiple starbursts, W($H\beta$) served as a good age indicator out to $\sim$10 Myr. 

Applying model-based age diagnostics to observations of star-forming galaxies requires accommodating for several complicating factors. Underlying continuum absorption from the lower-mass members of the young stellar population will decrease the observed value of the W($H\beta$) equivalent widths, making calibrations based only on the nebular emission features uncertain by as much as a factor of 2 at later ages and weaker nebular equivalent widths (Fernandes et al.\ 2003). Gonz\'{a}lez-Delgado (1999, 2005) present detailed examinations of how equivalent widths for the H$\alpha$ and H$\beta$ absorption features evolve with time for a starburst. Effectively accounting for this absorption component is crucial when developing a calibration meant to be applied to observations. Furthermore, a pure starburst model lacks any contribution from an older stellar population, which yields a much redder continuum and disproportionately dilutes observed equivalent widths for lines at shorter wavelengths; for further discussion of dilution effects see Fernandes et al.\ (2003) and Section 4. Finally, while stellar mass loss rates were found to be a secondary effect by Copetti et al.\ (1986), subsequent generations of stellar evolutionary models have newly emphasized the critical role that mass loss and other physical properties such as rotation and binarity can have on the lifetimes of stellar populations and even the particular phases predicted at a given mass by evolutionary tracks (e.g. Meynet et al.\ 1994, Vanbeveren et al.\ 1998Eldridge \& Stanway 2009, Ekstr\"{o}m et al.\ 2012, Levesque et al.\ 2012, Georgy et al.\ 2013). 

Here we present our theoretical relations between W($H\beta$), W(H$\alpha$), W(Br$\gamma$), and age. Our results are based on the latest generation of stellar evolutionary tracks and the Starburst99 stellar population synthesis code, and are the first to account for the combined effects of nebular emission {\it and} continuum absorption in our measurements of equivalent widths. In Section 2 we present the parameters adopted in our stellar population synthesis models and describe the analyses used to calculate our final equivalent widths. In Section 3 we illustrate the age evolution of W($H\beta$), W(H$\alpha$), W(Br$\gamma$) in our models and consider the effects produced by different star formation histories, variations to the IMF, and treatments of stellar rotation. Finally, in Section 4 we discuss persisting challenges for age diagnostics in observed spectra of star-forming galaxies as well as future improvements and analyses that could improve these diagnostics.

\section{Model Grid Parameters}
For our analysis of the W(H$\beta$), W(H$\alpha$), and W(Br$\gamma$) evolution with age, we use models produced by the Starburst99 evolutionary synthesis code (Leitherer et al.\ 1999, 2010; Leitherer \& Chen 2009; V\'{a}zquez \& Leitherer 2005). The Starburst99 code initially generates the mass distribution of a stellar population for a given metallicity $Z$ and following a specified initial mass function (IMF). Stellar evolutionary tracks are then used to assign fundamental stellar parameters (e.g. luminosity, effective temperature, etc.) and their variation with time commencing on the zero-age main-sequence and terminating when the end-point of stellar evolution is reached. For each time step during the evolution and for each mass interval, the code assigns properties such as the resulting energy distribution, color, chemical yields, and others. Integration over all masses and all stellar generations then provides the synthetic properties of the full population. The adjustable input parameters of the models are the initial chemical composition, the IMF as specified via the slope $\alpha$ and the upper and lower mass limits, and the star formation history. The stellar evolution models are fixed and considered well calibrated. It is important to realize that evolution models for massive stars are not derived from first principles but have built-in adjustable parameters, such as the mixing length and the mass loss rates, which were previously calibrated via observations of resolved stellar populations in the Local Group of galaxies. The main constraints are provided by the width of the main-sequence, ratios of various stellar types, and determinations of individual stellar parameters (e.g. Massey 2013). In order to produce realistic equivalent widths for H$\beta$ and H$\alpha$, we require models that account for the combined effects of the nebular emission and continuum as well as underlying stellar absorption, which becomes significant at 6-8 Myr in the optical regime (see Gonz\'{a}lez Delgado et al.\ 2005). We therefore use optical high-resolution output spectra from Starburst99 produced using the latest improvements in stellar atmosphere models, accommodating key components such as non-LTE effects and line blanketing (for a detailed discussion of these high-resolution models, see Martins et al.\ 2005). 

Starburst99 computes the ionizing fluxes from the set of model atmospheres published by Smith et al.\ (2002), which are assigned to each data point provided by the evolutionary tracks. The atmospheres are based on the WM-Basic (Pauldrach et al.\ 2001) and CMFGEN (Hillier \& Miller 1998) codes, which are generally considered the most appropriate for massive O- and Wolf-Rayet (WR) stars, respectively. Both sets of atmospheres are spherically extended and account for the hydrodynamics of the outflow (see the discussion in Smith et al.\ 2002 for a comparison of their main features). Both models sets were tailored by Smith et al.\ (2002) to match the properties of the Geneva evolutionary tracks and are therefore directly compatible with these evolution models. The WM-Basic O-stars models are used along the main-sequence and early post-main-sequence until the surface hydrogen abundance by mass reaches 0.4. Stars with lower hydrogen abundance are considered WR stars and are modeled with the CMFGEN atmospheres. The surface abundances of the major elements change drastically during the WR phase, and the CMFGEN atmosphere abundances are modified accordingly. While there is consistency between the CMFGEN abundances and the chemical composition of the tracks, this is not the case for the WM-Basic O-star models. In particular, the decrease of the hydrogen abundance is not accounted for. Fortunately, Starburst99 allows us to investigate whether the variable hydrogen abundance along the main-sequence tracks affect the emergent Lyman continuum. The Starburst99 code incorporates an additional set of WM-Basic models as part of a high-resolution UV library (Leitherer et al.\ 2010). These models cover the spectral region 900-3000\AA\ and do in fact account for the hydrogen and helium variation along the main-sequence. Comparison of the Smith et al.\ (2002) and Leitherer et al.\ (2010) models shortward of the Lyman edge suggest negligible differences.

The emergent fluxes of spherical, expanding atmospheres are sensitive to the wind densities, and therefore to the mass-loss rates. This sensitivity is particularly pronounced at the shortest wavelengths in the ionized helium continuum below 228\AA. The mass-loss rates of the atmosphere models adopted by Smith et al.\ (2002) are calibrated via observed values. In contrast, the mass-loss rates in the stellar evolution models are initially based on observations but then are adjusted to enforce agreement between the observed and theoretical upper Hertzsprung-Russell diagram. Generally, the rates in the evolution models are higher than those used for the atmospheres. However, this is not a concern for the hydrogen Lyman continuum, for which wind effects are almost negligible. This can be appreciated in Fig. 3 of Smith et al.\ (2002) where the UV part of the spectrum of a WM-Basic model is compared to a static plane-parallel Kurucz model (Lejeune et al.\ 1997). The latter can be thought of an atmosphere with zero mass loss. As expected, there are large differences below 228\AA, whereas the two models are identical (except for line-blanketing) in the hydrogen Lyman continuum. 

We include seven different sets of stellar evolutionary tracks in our models spanning six different metallicities. We begin with the set of non-rotating stellar evolutionary tracks presented in Meynet et al.\ (1994) that adopt a ``high" mass loss rate enhanced to approximate rotation effects. These tracks are calculated at metallicities of $Z=0.001, 0.004, 0.008, 0.020$, and 0.040 (here we define $Z$ as the mass-weighted heavy element abundance where the Sun has $Z=0.014$ on the scale discussed by Asplund et al.\ 2005). Note that the high-resolution stellar spectra in Starburst99 are available at metallicities of $Z=0.001, 0.008, 0.020$, and 0.040 (lacking $Z=0.004$); as a result, the $Z=0.004$ evolutionary tracks are paired with the $Z=0.008$ spectra. We also include the two new sets of stellar evolutionary tracks at $Z=0.014$ presented in Ekstr\"{o}m et al.\ (2012). Both included updated opacities, nuclear reaction rates, and mass-loss treatments as compared to the previous generation of models. Most importantly, one set of the new tracks models the evolution of non-rotating stars (initial $v_{\rm rot} = 0.0v_{\rm crit}$), while the other models the evolution of stars with an initial rotation velocity of $v_{\rm rot} = 0.4v_{\rm crit}$ (where $v_{\rm crit}$ is the break-up speed at the equator). For more discussion of these evolutionary tracks, see Ekstr\"{o}m et al.\ (2012). 

We adopt two different treatments of the SFH in our models: a zero-age instantaneous burst of star formation with a fixed mass of 10$^6 M_{\odot}$, and continuous star formation with a constant rate of 1 $M_{\odot}$ yr$^{-1}$, starting from an initial time of 0 Myr and assuming a stellar population large enough to fully sample the upper end of the IMF. Modeling an instantaneous burst of star formation allows us to clearly trace the effects of a single coeval stellar population, while a continuous star formation treatment represents the other extreme of a stellar population that reaches and sustains equilibrium after the first 3-4 Myr. In reality, most star-forming galaxies occupy a middle ground between the two extreme star formation histories that we model here, showing evidence of several short bursts of star formation interleaved with quiescent periods (see, for example, Terlevich et al.\ 2004, Mart\'{i}n-Manj\'{o}n et al.\ 2008). Starting with an initial 0 Myr age, we simulate the evolution of our synthetic stellar populations up to 20 Myr in 0.5 Myr increments. At ages $>$20 Myr for the instantaneous burst stellar populations, the galaxy spectrum becomes dominated by lower-mass stars and the nebular emission features disappear as the hot ionizing massive star population dies out.

Our initial grid of Starburst99 outputs was run with a Kroupa IMF, with $\alpha=1.3$ for the 0.1M$_{\odot}$-0.5M$_{\odot}$ mass range and $\alpha=2.3$ for 0.5M$_{\odot}$-100M$_{\odot}$ (Kroupa 2001). However, in order to explore the effects of the IMF on our age diagnostic lines, we also ran models for the $Z=0.014$ rotating and non-rotating stellar populations that adopt $\alpha=1.3$ and $\alpha=3.3$ for the 0.5M$_{\odot}$-100M$_{\odot}$ mass range (the regime of interest for our work given the dominance of massive stars at these early ages) to examine the impact of varying IMFs. All three IMFs have a mass range of 0.1$M_{\odot}$-100M$_{\odot}$.

The combined nebular emission and continuum components of W(H$\beta$), W(H$\alpha$), and W(Br$\gamma$) were determined directly from the \texttt{ewidth} output file of the Starburst99 code, which includes the results of the code's determinations of continuum luminosities, line luminosities, and equivalent widths for the nebular hydrogen features (for more discussion on the importance of the nebular continuum, particularly at early ages and longer wavelengths, see Leitherer \& Heckman 1995). The stellar absorption components were calculated based on the \texttt{hires} Starburst99 output file. These data include the code's calculation of a high-resolution theoretical spectrum from 3000-7000\AA \footnotemark. For our calculations we used the normalized spectra from this file to fit the stellar H$\beta$ and H$\alpha$ absorption features and determine an equivalent width. Combining these (negative) values of absorption equivalent widths with the nebular components at each timestep produced final values of W(H$\beta$) and W(H$\alpha$) for our models. This correction for stellar absorption is particularly important in the case of our instantaneous burst star formation history models at later ages, where the nebular emission components are depleted and the stellar absorption increases (for more discussion see Gonz\'{a}lez Delgado et al.\ 2005). In the case of Br$\gamma$, the stellar absorption in this wavelength regime is insignificant at early ages; the continuum is dominated by red supergiants, which do not typically show hydrogen in their spectra (see, for example, Levesque et al.\ 2005), at later ages ($\gtrsim$10 Myr). As a result we simply calculate W(Br$\gamma$) as the nebular components with no additional contribution from stellar absorption.

\footnotetext{For more information on the output products of Starburst99 please see \texttt{http://www.stsci.edu/science/starburst99/}.}

\section{Evolution of the Hydrogen Recombination Lines}
Figure 1 shows the evolution of W(H$\beta$), W(H$\alpha$), and W(Br$\gamma$) with time for our models adopting an instantaneous burst star formation history. All three spectral features show a similar evolution with time and a metallicity-driven spread, comparable to the results of Schaerer \& Vacca (1998) and Leitherer et al.\ (1999). Their evolution is best characterized by a sharp drop in the first 2-5 Myr followed by a significantly more graduate decrease to a $\le$0 equilibrium state. The ages at which these changes take place show a clear dependence on metallicity: equivalent widths in the non-rotating $Z = 0.014, 0.020$ and $0.040$ models show a rapid decrease beginning at 0 Myr and diminish to $\le$0 by 6.5 Myr, while at lower metallicities (and solar metallicity in the case of the rotating models) the equivalent widths stay nearly constant for the first 1-2 Myr and don't drop to $\le$0 until much later. The 0-2Myr plateaus produced by those models can be attributed to longer main-sequence lifetimes for massive stars, allowing these stars to remain in a hotter evolutionary state for longer in such conditions (a consequence of weaker line-driven mass-loss in the low-metallicity models and mixing-driven refueling of the hydrogen-burning core in the rotating models; Ekstr\"{o}m et al.\ 2012). The clearest difference in the later-age evolution is apparent in the $Z = 0.001$ models, which shows a persistent emission component in all three line features that is present up to $\sim$15 Myr for W(H$\beta$) and W(H$\alpha$) and $\sim$18 Myr for W(Br$\gamma$). 

Our models with $Z = 0.004, 0.008$, and 0.014 also show a brief plateau or increase in W(H$\beta$), W(H$\alpha$), and W(Br$\gamma$) beginning at 3-3.5 Myr and lasting 0.5-1Myr (interestingly, the evolution of W(H$\beta$), W(H$\alpha$), and W(Br$\gamma$) illustrated in Leitherer et al.\ 1999 also shows this small increase, but only in the {\it higher}-metallicity cases). The rotating $Z = 0.014$ models include a second 0.5-1Myr increase in the equivalent widths at 5.5 Myr, and predict larger equivalent widths than the non-rotating $Z = 0.014$ models between 1-10.5 Myr. From our past work on the ionizing spectra produced by rotating and non-rotating stellar populations (Levesque et al.\ 2012) we can attribute these small increases in equivalent width to the onset and duration of the Wolf-Rayet phase in our modeled stellar populations. Non-rotating models of stellar evolution predict Wolf-Rayet stars an age of 3-5 Myr, while rotating models permit lower-mass stars to evolve to the Wolf-Rayet phase, maintaining a higher ionizing flux for longer and showing a later increase in equivalent widths corresponding to the age of the lower-mass Wolf-Rayet population.

The exception to this very similar evolution of the hydrogen recombination lines with time is the $Z = 0.001$ model. At this metallicity we see a slower overall decrease in W(H$\beta$) and W(H$\alpha$) with time and a significant resurgence in the value of W(Br$\gamma$) beginning at 5.5 Myr and lasting until 7 Myr. This demonstrates that Br$\gamma$ may not be a robust indicator of young stellar population age at very low metallicities due to its double-valued nature of behavior. The unique behavior at $Z=0.001$ can be attributed to the effects of low metallicity on the red supergiant population. At high metallicities the ionizing fluxes decrease nearly monotonically after 3 Myr and the near-IR continuum increases at $>$5 Myr due to the onset of the red supergiant phase for moderately-massive (10-25M$_{\odot}$) stars. Combined, this leads to a steep decline in equivalent width. At $Z=0.001$, the behavior of the ionizing fluxes is the same but there are far fewer red supergiants, a consequence of low-metallicity stellar evolution (see, for example, Elias et al.\ 1985, Levesque \& Massey 2012). This lack of red supergiants causes a significant drop in the near-IR continuum luminosity at 5-7 Myr, leading to an ultimate increase in the observed equivalent width at these ages (see also Leitherer et al.\ 1999). A similar but much slighter ``bump" is also apparent in the $Z=0.004$ models at 5.5 Myr. It is worth noting that, as described in Section 2, our $Z = 0.004$ models were produced by combining $Z = 0.004$ stellar evolutionary tracks with $Z = 0.008$ atmosphere models; therefore, it is possible that a model run with $Z = 0.004$ stellar atmospheres would in fact present a clearer missing link between the predictions of our $Z = 0.001$ and $Z = 0.004$ models.

In Figure 2, we plot W(H$\beta$), W(H$\alpha$), and W(Br$\gamma$) as a function of time for models that adopt a continuous star formation rate. While our instantaneous burst models are the best means of examining the equivalent widths produced by a single coeval stellar population - making them the preferred models for applications for starburst galaxies - our continuous star formation models represents the opposing ``extreme" of star formation history. With all star-forming galaxies occupying some middle ground between these two assumptions, the equivalent widths produced by both sets of models effectively serve as lower (instantaneous) and upper (continuous) limits on the young stellar population age associated with an observed equivalent width. Like the instantaneous burst models, our continuous star formation models show a similar metallicity-dependent evolution with time for W(H$\beta$), W(H$\alpha$), and W(Br$\gamma$); however, in the continuous star formation case the models instead approach constant $>$0 values at later ages as the synthetic stellar populations reach equilibrium. At later ages in particular, these predicted equivalent widths are useful for corrections in broad-band photometry observations of star-forming galaxies.

Finally, in Figure 3 we consider the effects of IMF variations on the predicted values of W(H$\beta$), W(H$\alpha$), and W(Br$\gamma$) for our rotating and non-rotating $Z = 0.014$ models. As previously illustrated by Leitherer et al.\ (1999), changes in $\alpha$ can have a strong effect on the predicted equivalent widths of the hydrogen recombination lines; steeper IMFs consistently produce smaller equivalent widths at all ages and the difference appears to become more pronounced at higher values of $\alpha$. A Kroupa IMF and shallower ($\alpha = 1.3$ throughout) IMF do produce comparable values of W(H$\beta$), W(H$\alpha$), and W(Br$\gamma$) for the non-rotating instantaneous burst models, but otherwise the differences are consistent for both star formation histories and rotation treatments. Based on a similar analysis in Leitherer et al.\ (1999), these IMF effects should not be metallicity-dependent. 

\section{Discussion}
We have illustrated the predicted evolution of W($H\beta$), W(H$\alpha$), W(Br$\gamma$) as a function of age for our grid of stellar population synthesis models. These predictions are the first to accommodate the combined effects of nebular emission and continuum absorption produced by a stellar population, and also include models that span multiple metallicities, star formation histories, and stellar rotation treatments, offering improvements over previous work (e.g. Leitherer et al.\ 1999, Schaerer \& Vacca 1998, Mart\'{i}n-Manj\'{o}n et al.\ 2008, Levesque et al.\ 2010a). Given constraints on metallicity, these predicted equivalent widths can be applied to observations of star-forming galaxies to approximate the age of their young stellar populations.

However, it is important to note several shortcomings of these models that may limit the efficacy of our predicted equivalent widths when applied to observations. Like many stellar population synthesis codes (e.g. Bruzual \& Charlot 2003), Starburst99 does not account for the destructive effects of dust, instead assuming that 100\% of the ionizing photons produced by the stellar population make it to recombination. Accounting for dust effects in typical star-forming galaxies should yield values for W(H$\alpha$) and W(H$\beta$) that are $\sim$30\% lower to account for the absorption of Lyman continuum photons (e.g. DeGioia-Eastwood 1992, Fioc \& Rocca-Volmerage 1997, Inoue et al.\ 2001). Similarly, the use of hydrogen recombination lines as age indicators is rendered ineffective for very high-metallicity or dusty galaxies where the bulk of the ionizing photons will be destroyed before they can recombine.

The predicted W(H$\alpha$) and W(H$\beta$) equivalent widths can also potentially suffer from the effects of dilution, as described in Fernandes et al.\ (2003) and Section 1. The presence of a significantly older ($\gtrsim$1 Gyr) stellar population in a star-forming galaxy will result in a significantly redder continuum (due to domination by asymptotic giant branch stars) and significantly dilute the equivalent widths of optical emission features. While it is worth noting that such a scenario assumes a very specific and burst-like star formation history - multiple starbursts separated by $\ge$1 Gyr - many star-forming galaxies do include a significant older stellar population that will at least nominally impact the observed continuum. Mart\'{i}n-Manj\'{o}n et al.\ (2008) discuss these effects in more detail in their models of multi-burst star formation histories and propose combining emission line observations with $U-V$ colors for star-forming galaxies to effectively characterize the presence of older underlying stellar populations.

Finally, it is clear that treatments of stellar rotation and mass loss in the evolutionary tracks can have a significant effect on the predicted equivalent widths produced by stellar populations. Levesque et al.\ (2010b) highlighted the insufficient ionizing fluxes produced by Starburst99 when adopting the non-rotating stellar evolutionary tracks of Meynet et al.\ (1994). Subsequently, in Levesque et al.\ (2012) we examined the ionizing continuum produced by a rotating stellar population in detail, and conclude that the rotation rates modeled by the Geneva evolutionary tracks (Ekstr\"{o}m et al.\ 2012, Georgy 2012) may now lead to an overproduction of high-energy ionizing photons. The treatment of rotation itself presents an important source of uncertainty that must be considered in such work (e.g. Meynet et al.\ 2013, Chieffi \& Limongi 2013). Finally, current evolutionary tracks, both rotating and non-rotating, have highlighted several uncertainties concerning mass loss rates, particularly during the red supergiant phase (Georgy et al.\ 2012).  Even so, it is clear that rotation directly yields larger equivalent widths for the hydrogen recombination lines. Furthermore, binary stellar evolution also has a significant impact on the ionizing flux; similar to a rotation stellar population, models of binary stellar evolution produce significantly larger equivalent widths (Eldridge \& Stanway 2009). These comparisons highlight the importance of stellar evolutionary models, and their treatments of crucial properties such as rotation and binary, in modeling stellar populations. \\

EML is supported by NASA through Hubble Fellowship grant  number HST-HF-51324.01-A from the Space Telescope Science Institute, which is operated by the Association of Universities for Research in Astronomy, Incorporated, under NASA contract NAS5-26555. Support for Program number AR-12824 was provided by NASA through a grant from the Space Telescope Science Institute, which is operated by the Association of Universities for Research in Astronomy, Incorporated, under NASA contract NAS5-26555.

\begin{figure}
\epsscale{0.55}
\plotone{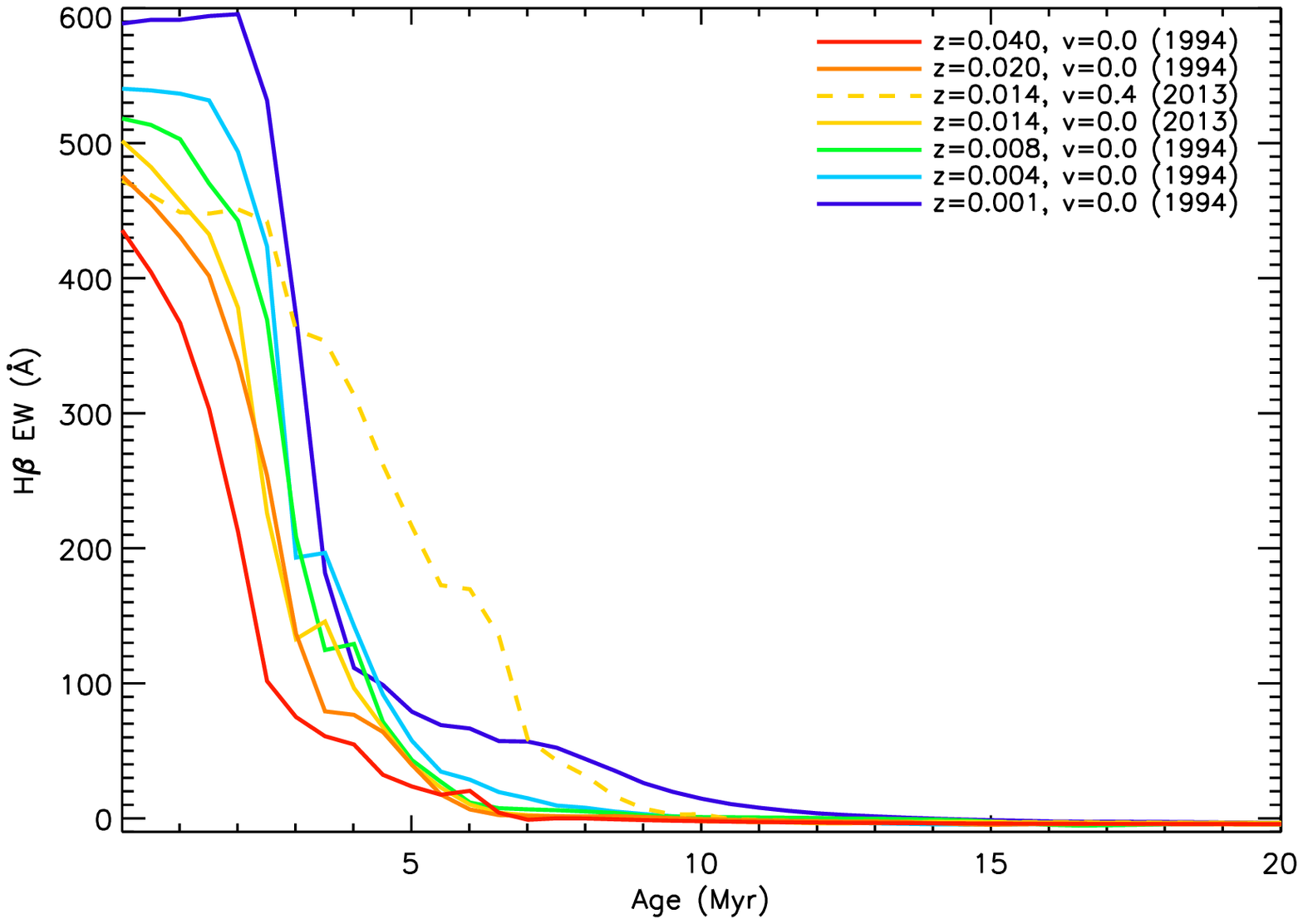}
\plotone{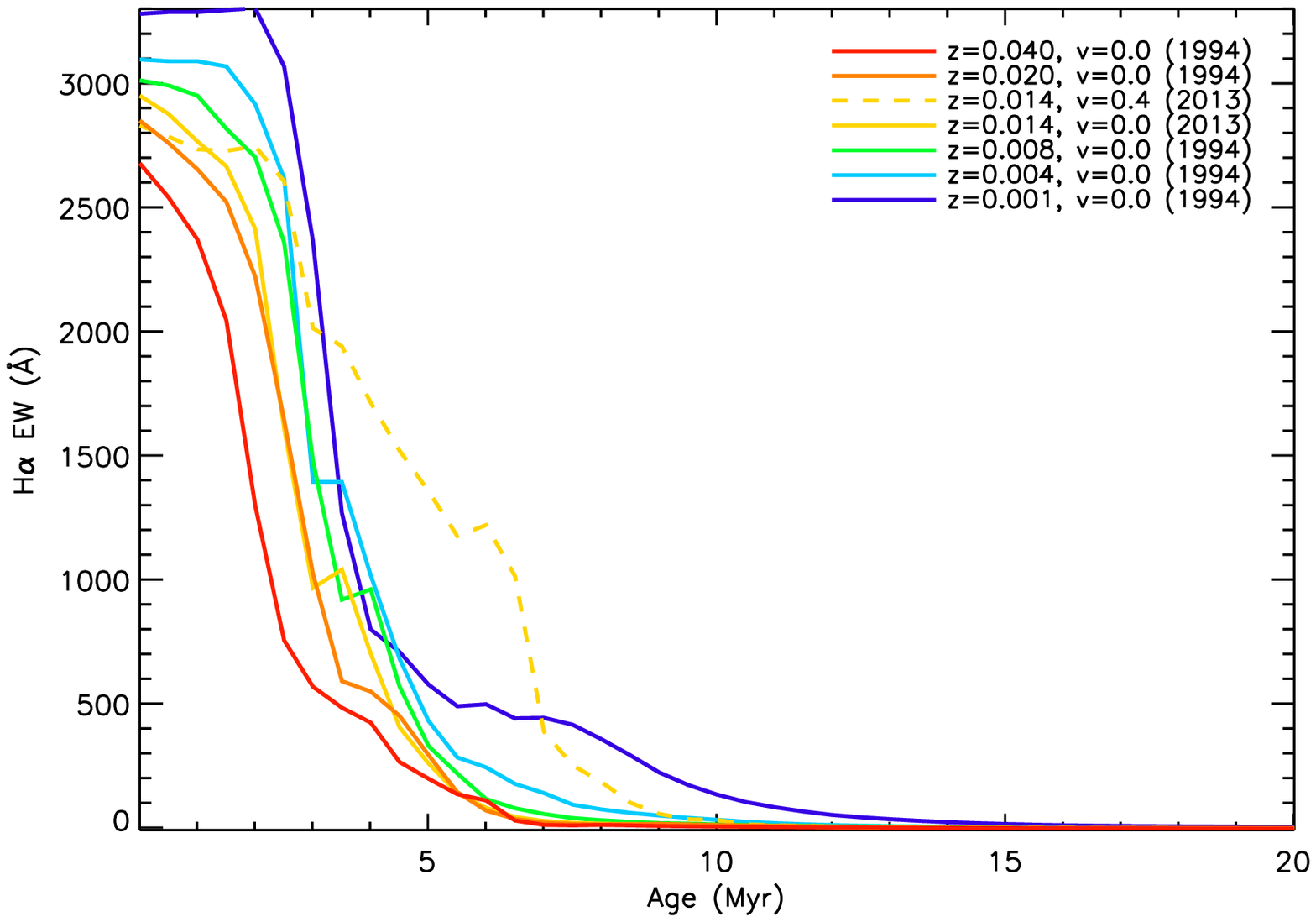}
\plotone{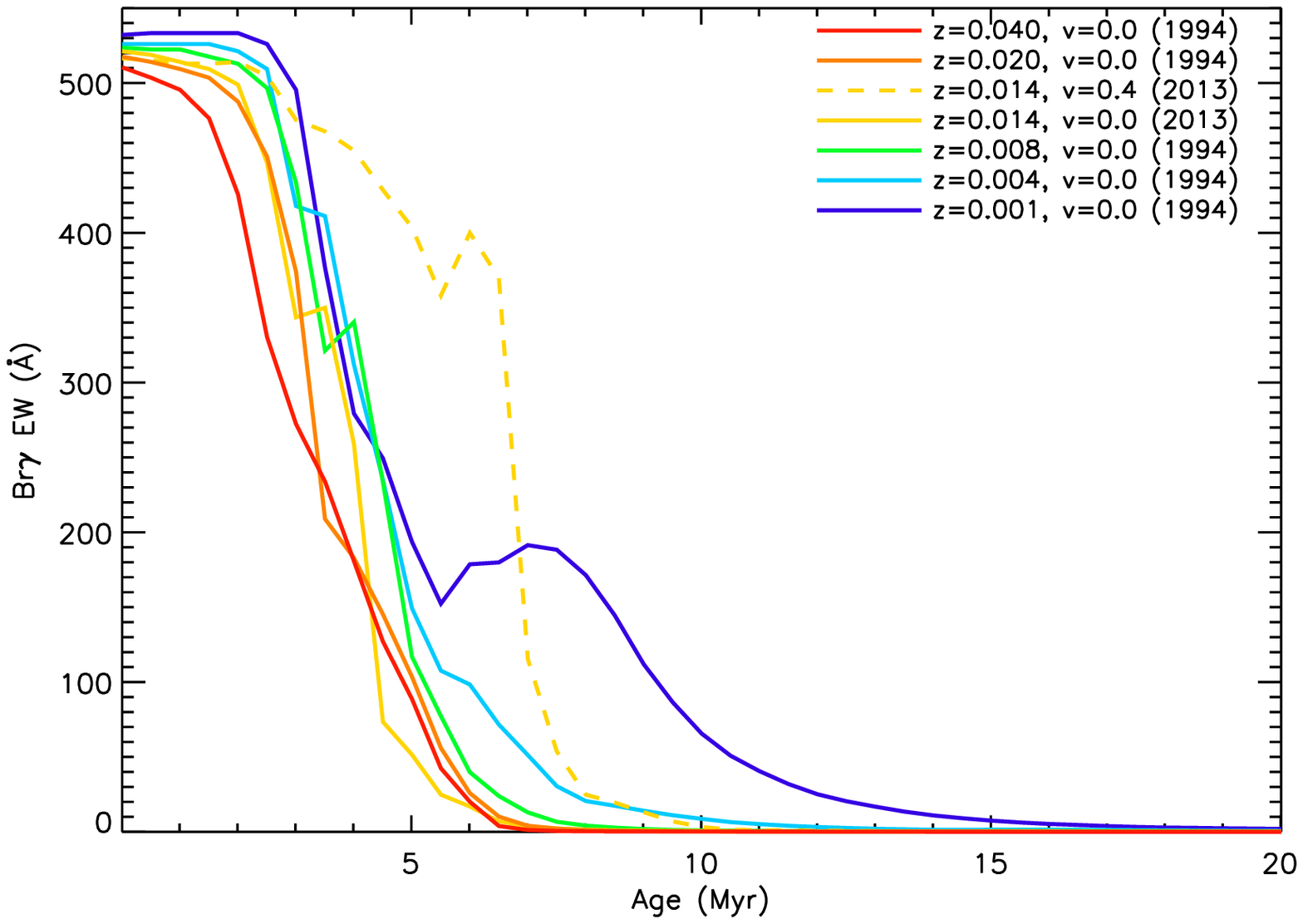}
\caption{The evolution of W(H$\beta$) (top), W(H$\alpha$) (center), and W(Br$\gamma$) (bottom) with age for our models adopting an instantaneous burst star formation history (with an initial mass of 10$^6$ M$_{\odot}$) and a Kroupa IMF ($alpha$ = 2.3 for 0.5M$_{\odot}$-100M$_{\odot}$). The models span the five metallicities available from the Meynet et al.\ (1994) stellar evolutionary tracks as well as the $z=0.014$ evolutionary tracks of Ekstr\"{o}m et al.\ (2012), and adopt rotation rates of $v_{\rm rot} = 0$ (solid lines) and $v_{\rm rot} = 0.4v_{\rm crit}$ (dashed line).}
\end{figure}

\begin{figure}
\epsscale{0.55}
\plotone{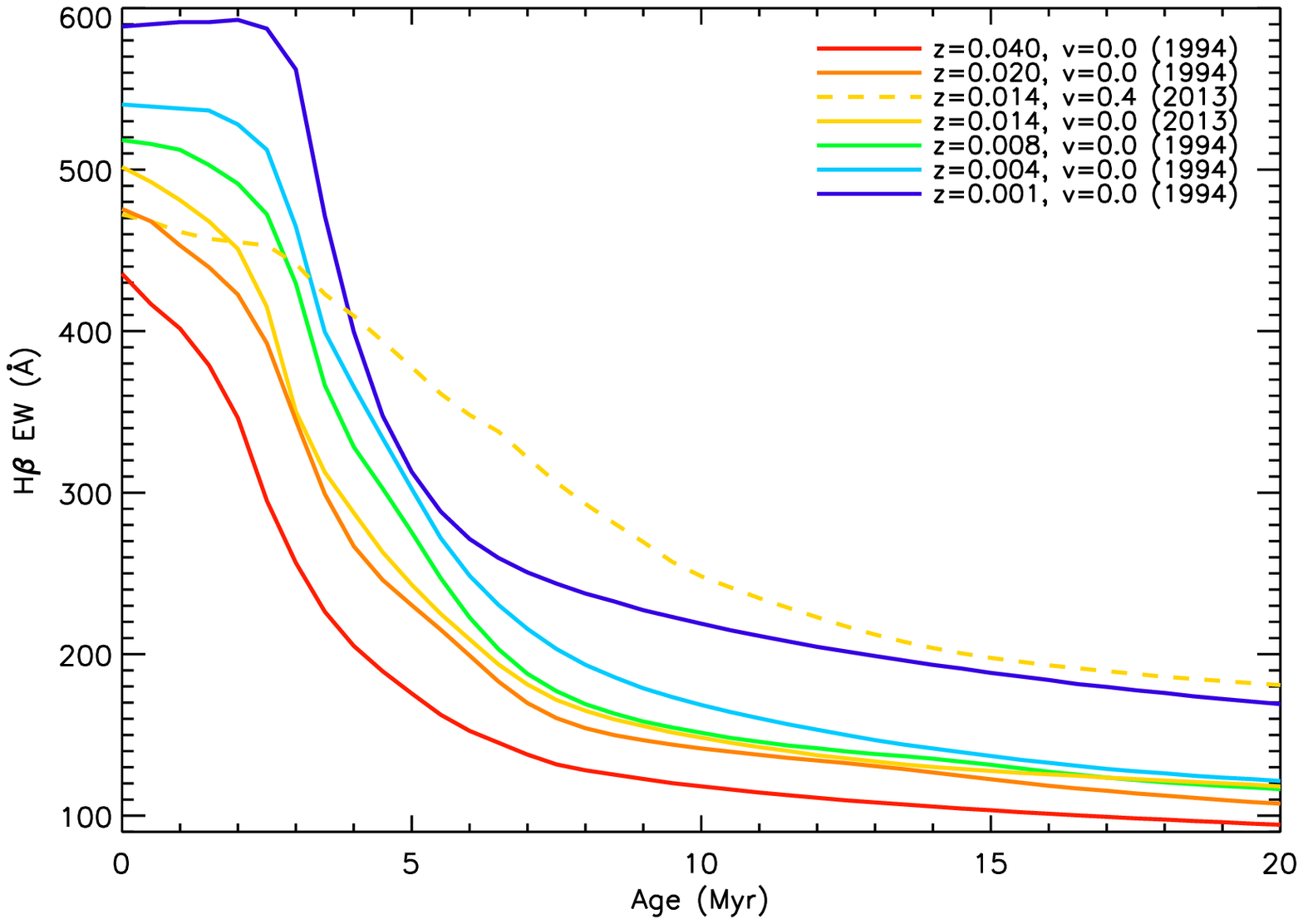}
\plotone{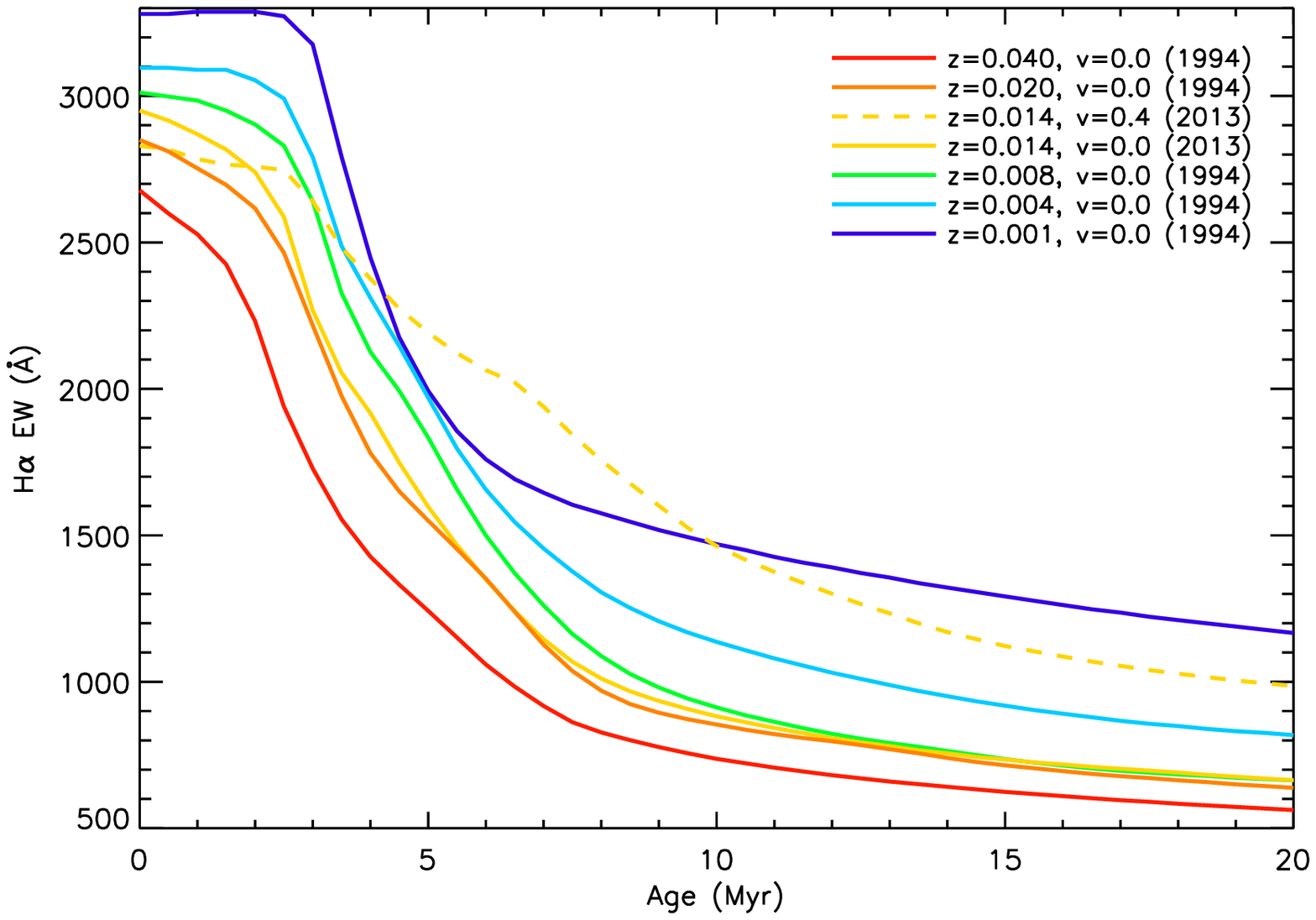}
\plotone{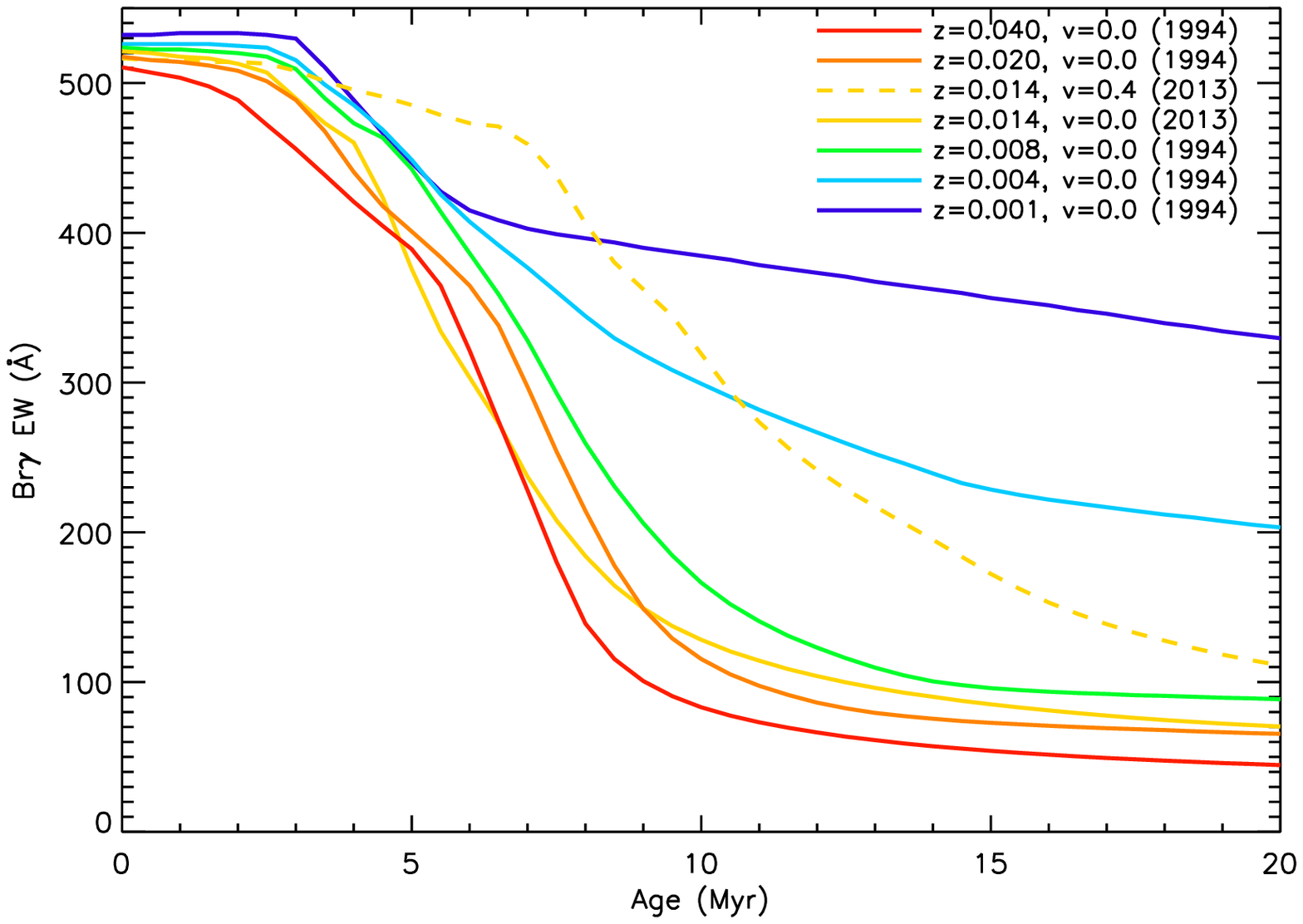}
\caption{As in Figure 1, but for models adopting a continuous star formation history with a star formation rate of 1 $M_{\odot}$ yr$^{-1}$.}
\end{figure}

\begin{figure}
\epsscale{0.49}
\plotone{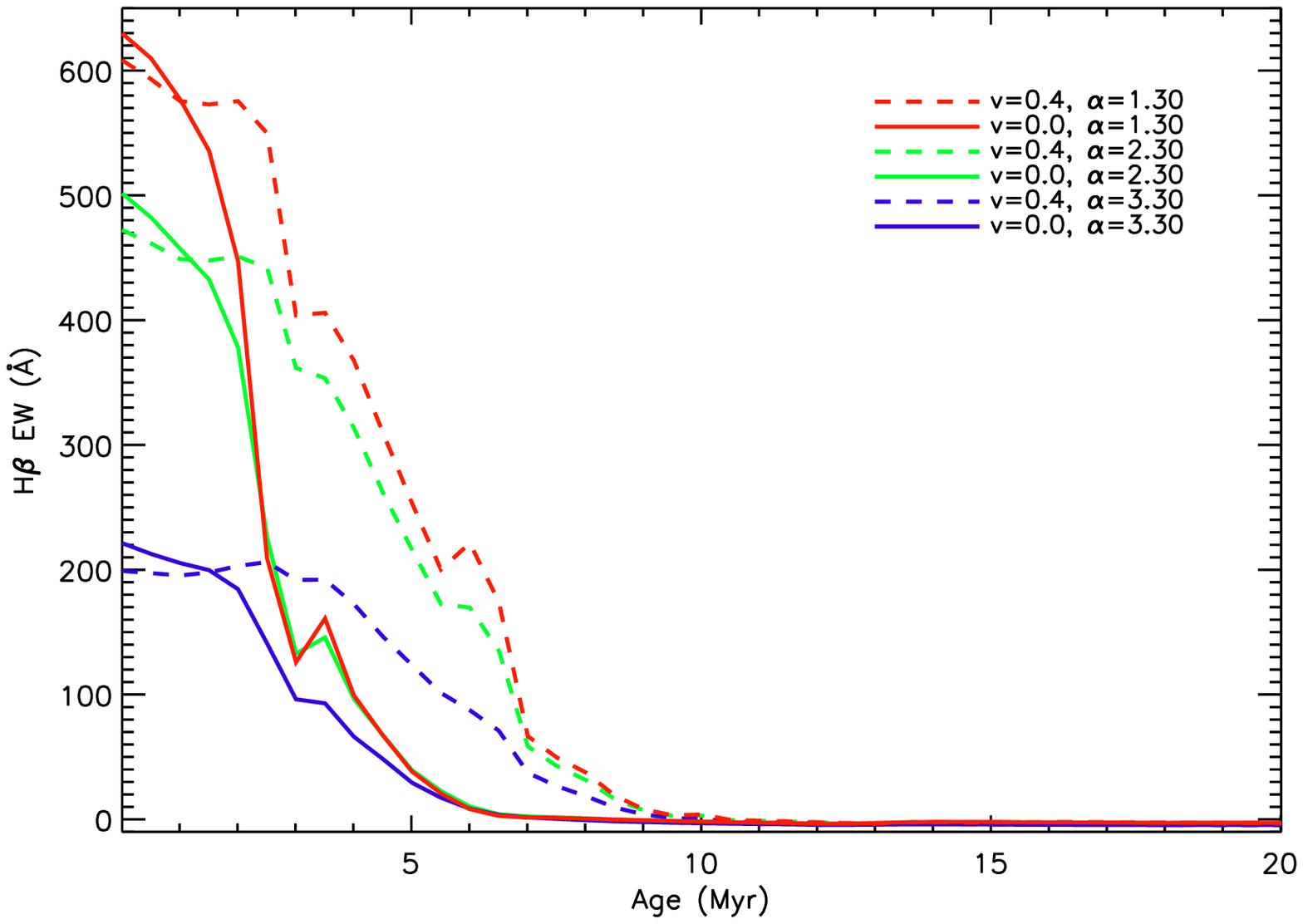}
\plotone{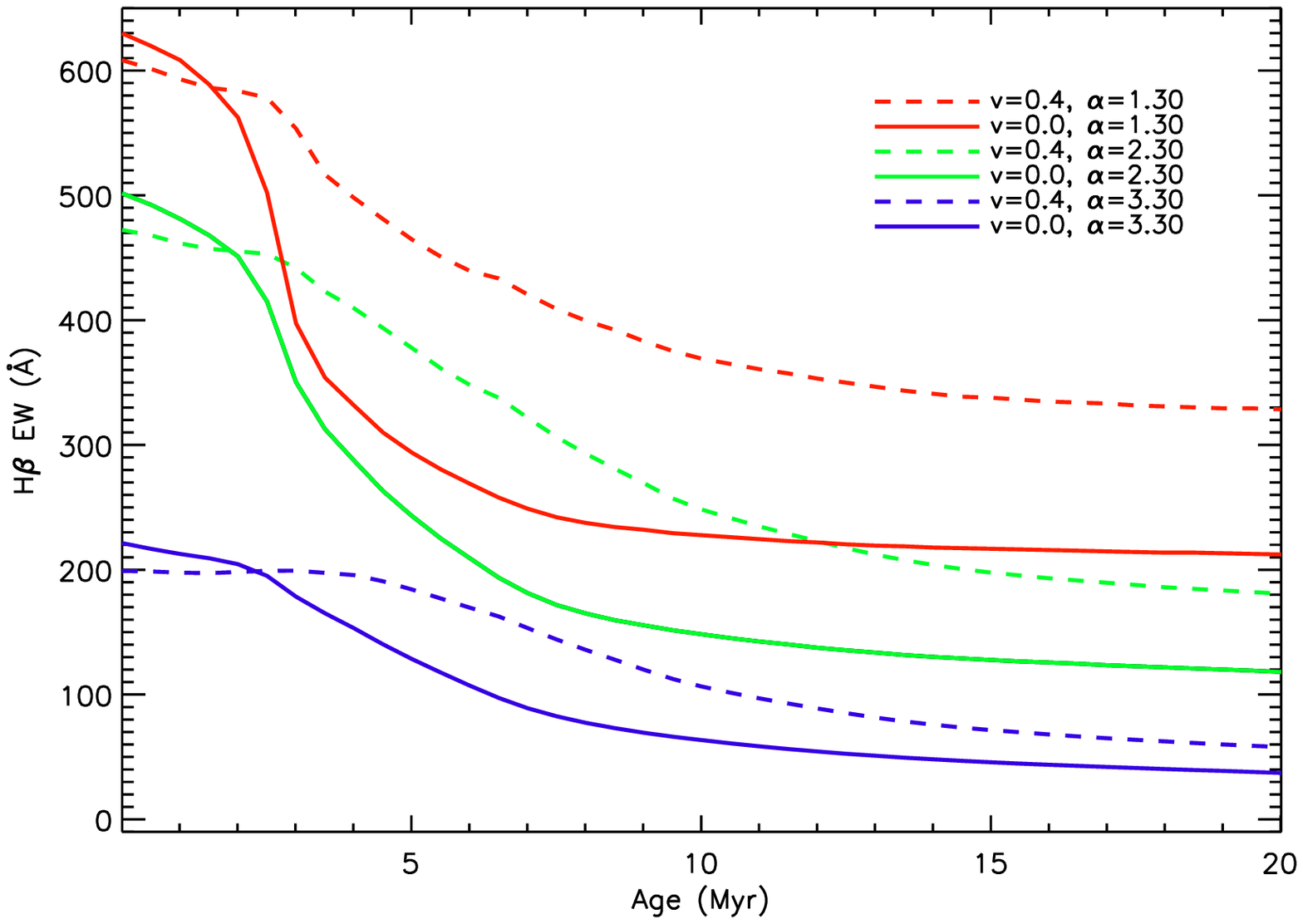}
\plotone{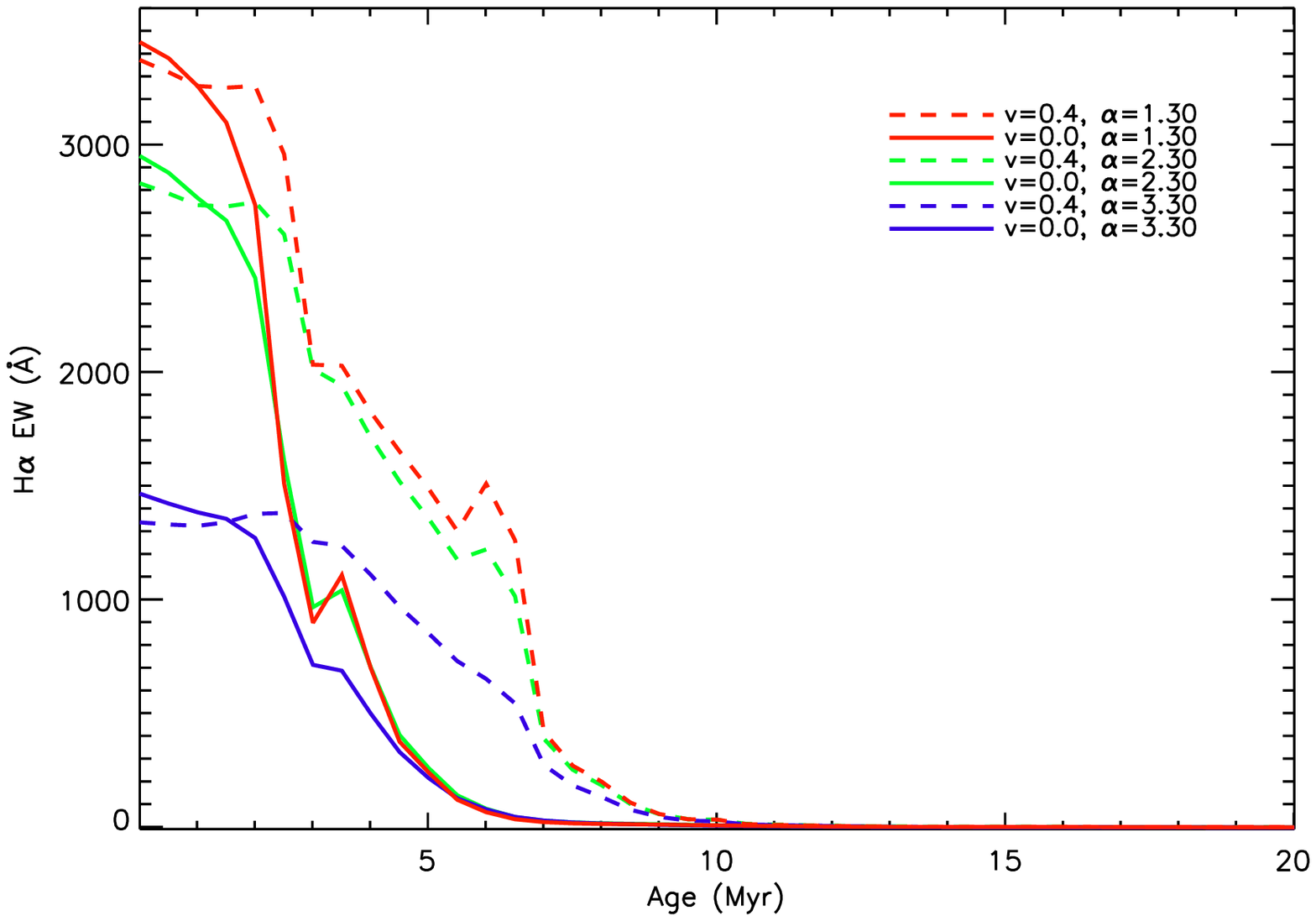}
\plotone{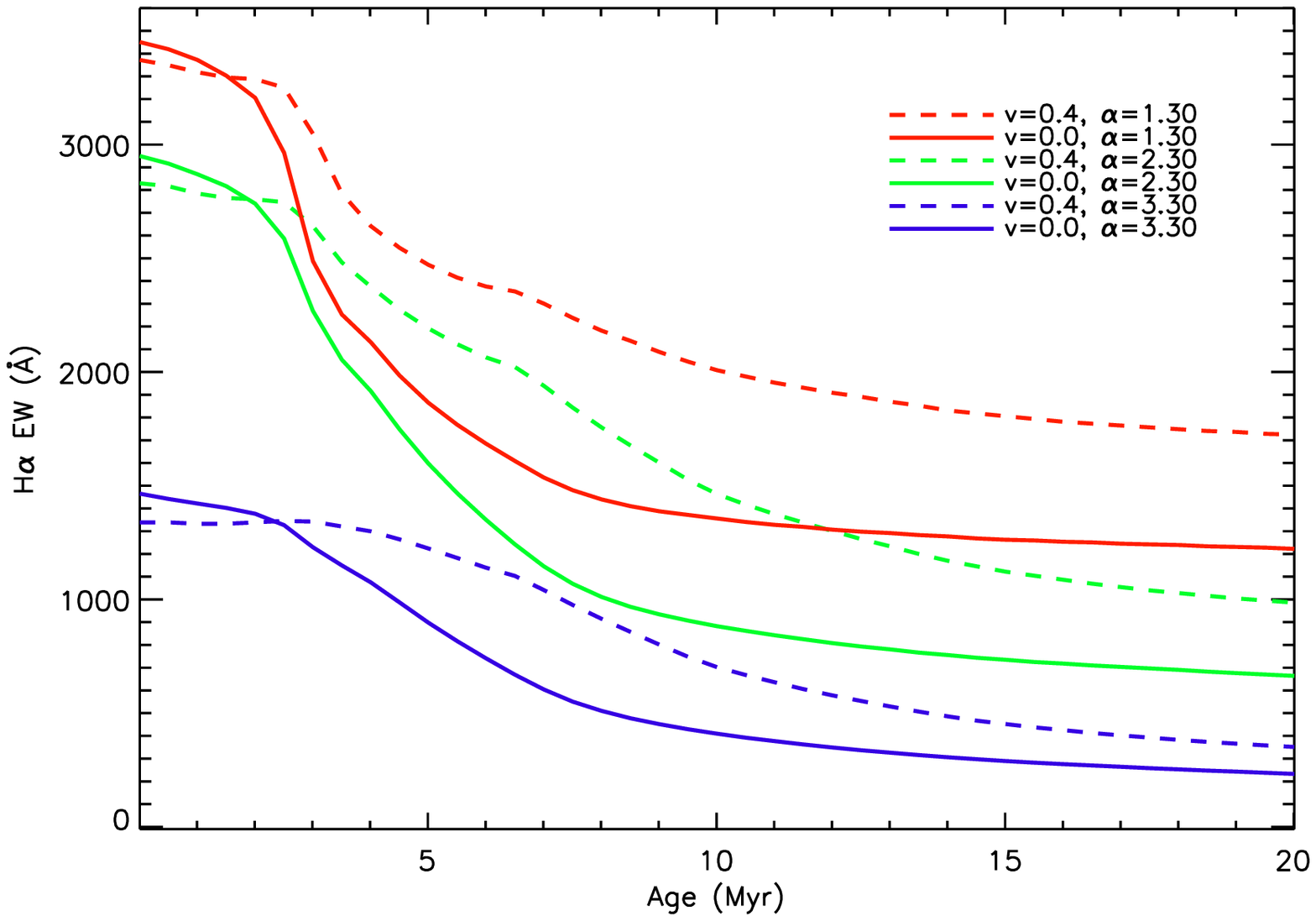}
\plotone{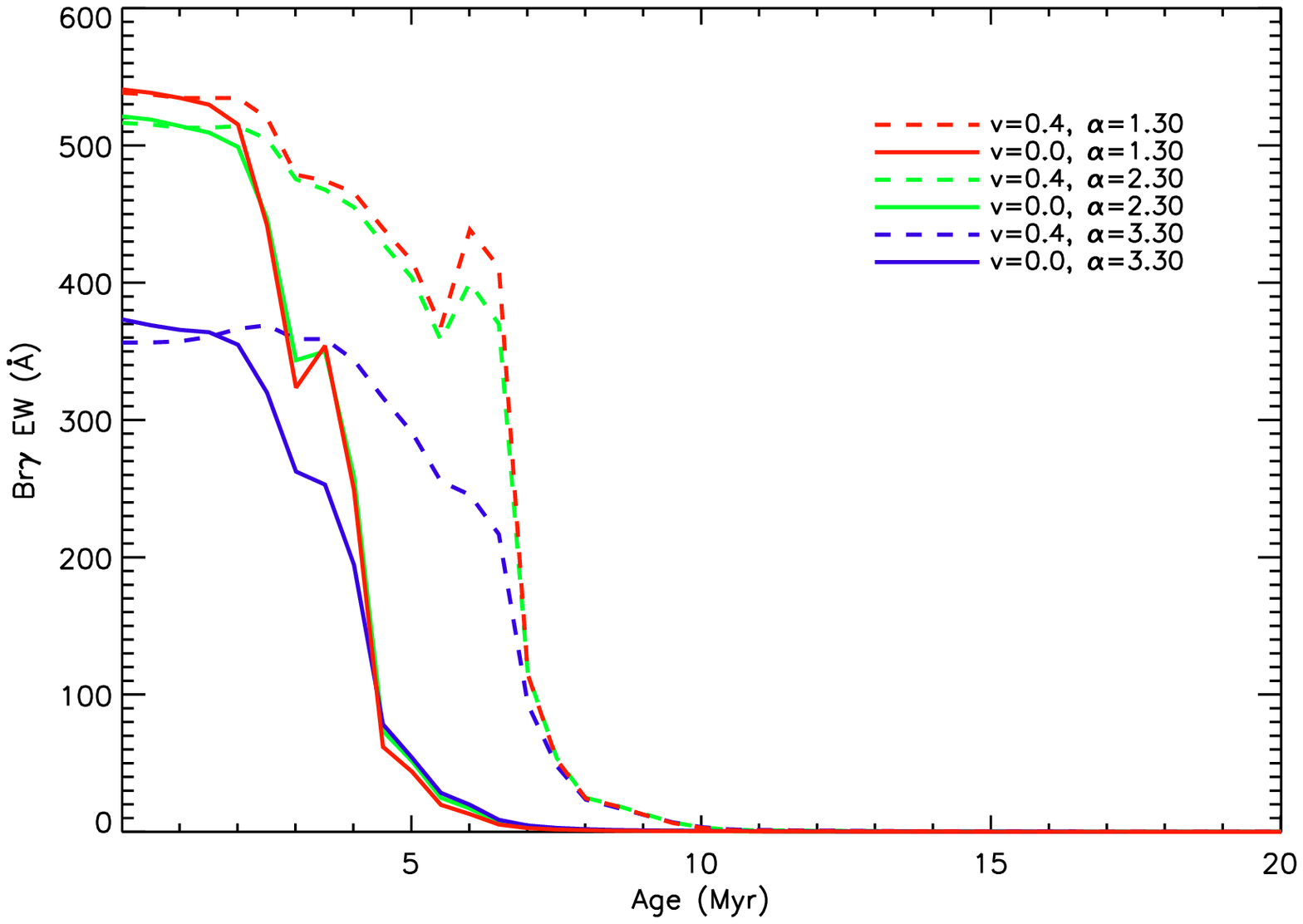}
\plotone{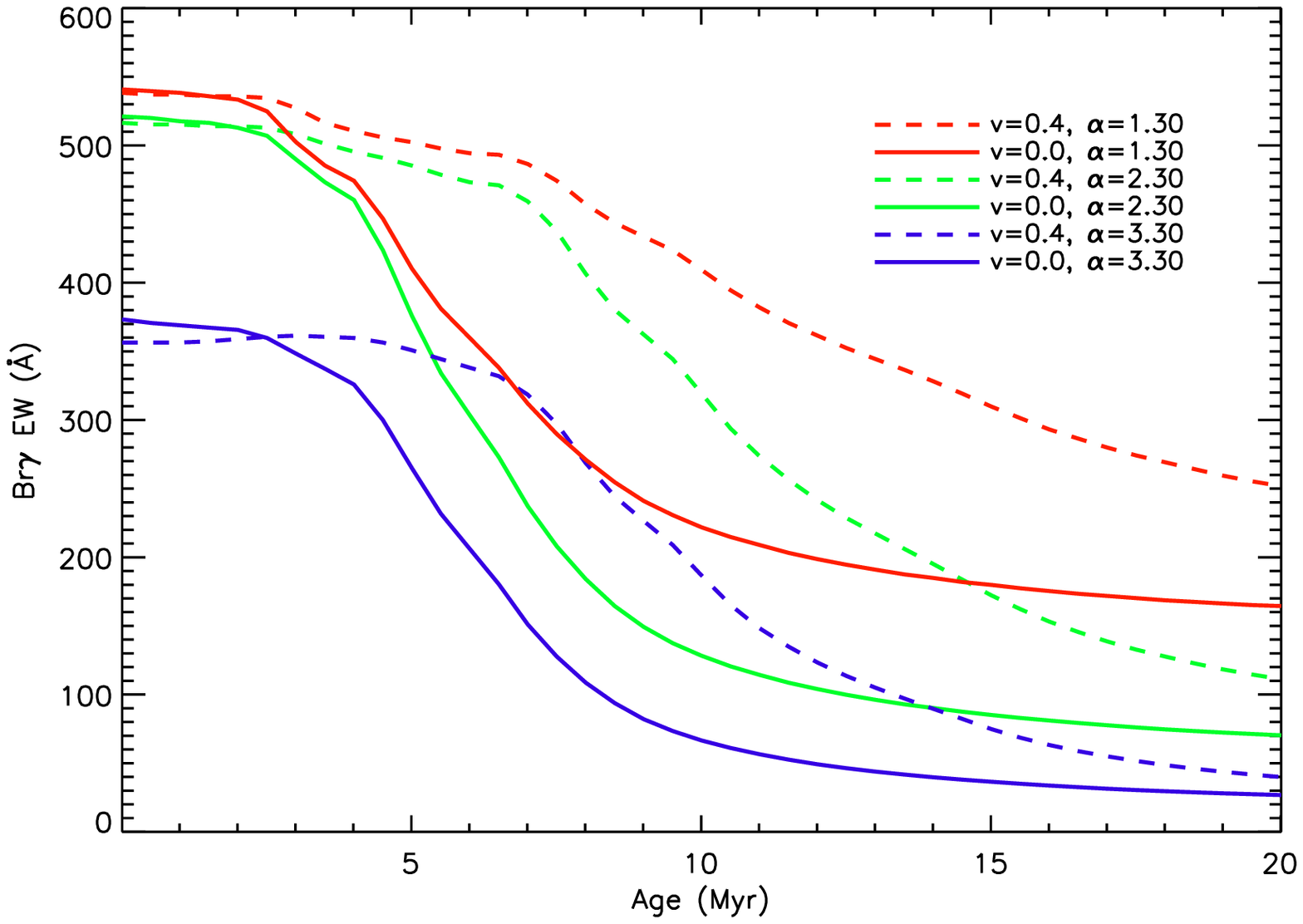}
\caption{The evolution of W(H$\beta$) (top), W(H$\alpha$) (center), and W(Br$\gamma$) (bottom) with age for our instantaneous burst (left) and continuous star formation (right) models, demonstrating the effects of varying the IMF. The models shown here adopt the $z=0.014$ $v_{\rm rot} = 0$ (solid lines) and $v_{\rm rot} = 0.4v_{\rm crit}$ (dashed lines) evolutionary tracks of Ekstr\"{o}m et al.\ (2012) and IMFs with a 0.5M$_{\odot}$-100M$_{\odot}$ mass range exponent of $\alpha=1.3$ (red), $\alpha = 2.3$ (Kroupa, green), and $\alpha = 3.3$ (blue).}
\end{figure}

\end{document}